\documentclass[12pt]{article}
\usepackage{oldgerm}
\usepackage{yfonts}
\usepackage{euler}
\usepackage{graphics}
\usepackage{bm}
\usepackage{graphicx}
\usepackage{amsmath}
\usepackage{amssymb}
\usepackage{amstext}
\usepackage{epstopdf}
\usepackage{amscd}
\usepackage{amsfonts}
\usepackage{appendix}
\usepackage{color}
\DeclareMathAlphabet{\EuFrak}{U}{euf}{m}{n}
\DeclareMathAlphabet{\EuScript}{U}{eus}{m}{n}

\newcommand{\nd}{\noindent}

\newcommand{\be}{\begin{equation}}
\newcommand{\ee}{\end{equation}}
\newcommand{\ben}{\begin{eqnarray}}
\newcommand{\een}{\end{eqnarray}}

\title{{\bf Analysis of Tsallis' classical partition function's poles  }}
\author{{A. Plastino$^1$, M. C. Rocca$^1$}\\,
\small{$^1$ La Plata National University and
Argentina's National Research Council}\\
\small{(IFLP-CCT-CONICET)-C. C. 727, 1900 La Plata - Argentina}}

\date{\today}

\begin{document}

\maketitle

\begin{abstract}

\nd When one integrates the q-exponential function of Tsallis' so as to get the partition function $Z$, a gamma function inevitably emerges. Consequently,  poles arise. We investigate here here the thermodynamic significance of  these poles in the case of $n$ classical harmonic oscillators (HO). Given that this is an exceedingly well known system, any new feature that may arise can safely be  attributed to the poles' effect.
  We appeal to the mathematical tools used in [EPJB 89, 150 (2016) and
 ArXiv:1702.03535 (2017)], and obtain both bound and unbound states. In the first case, we are then faced with a classical  Einstein crystal.
We also detect what might be interpreted as pseudo gravitational effects.

\nd Keywords: q-Statistics, divergences,  partition function,
dimensional regularization, specific heat.

\end{abstract}

\newpage

\renewcommand{\theequation}{\arabic{section}.\arabic{equation}}

\section{Introduction}

\setcounter{equation}{0}

\nd Tsallis'  q-statistical mechanics yielded
variegated applications in the last  25 years
\cite{tsallis,web,pre,epjb1,epjb2,epjb3,epjb4,epjb5,epjb6,
epjb7,epjb8,epjb9}. This statistics is of great
importance for  astrophysics, in what respects to self-gravitating systems \cite{PP93,chava,lb}.
 Further, it was shown to be useful in  diverse scientific fields. It has to its credit several
thousands  of  papers and authors \cite{web}. Investigating
its structural characteristics should be important for astronomy, physics,
neurology, biology, economic sciences, etc. \cite{tsallis}. Paradigmatic example is found in
its  application to high energy physics, where the q-statistics seems to describe well the
transverse momentum distributions of different hadrons
\cite{tp11o,tp11,phenix}.

\nd In this work we use standard mathematical tools
described in \cite{epjb, arxiv} to investigate interesting
properties of the Tsallis statistics of n harmonic
oscillators.

\nd The central point is the fact that the integrals
used to evaluate the partition function $Z$ and the
mean energy $<U>$ diverge for specific q-values.
These divergences can be overcome as described in
\cite{epjb, arxiv}

\nd A basic result to be obtained here is that the number of classical oscillators, $n$,
is strongly limited by the dimensionality $\nu$ and
the Tsallis parameter q. For $<U>>0$ and $Z>0$, i.e.,
the conventional theory,  $n$ must be finite and bounded.

\vskip 3mm

\nd A different panorama emerges by recourse to analytical extension in $\nu$. Then it is possible
to have a situation in which
$Z>0$, $<U><0$, $C<0$, with n finite and bounded.
Thus, our systems are here bound, representing a ''classical crystal'', and also self-gravitating
\cite{lb}. Finally, we will study the theory's poles
by recourse to dimensional regularization \cite{epjb, arxiv}.
We find at the poles, that i) the specific heat  $C$ is temperature ($T$) dependent (classically!), and,
ii) again, gravitational effects. Note the  $C$ can be $T-$dependent only due to internal degrees
of freedom, and that this is a quantum effect. We detect this dependence here at a purely classical level.

 \vskip 3mm
\nd We are motivated by the need of trying to determine what kind of hidden correlations are entailed by the non additivity of Tsallis' entropy $S_q$  for two independent systems A, B, i.e.,

$$S_q(A,B) = S_q(A) + S_q(B) + (1-q) S_q(A)S:q(B); \,\,\, q \in {\cal R}.$$
This is conveniently done by appeal to quite simple systems, whose physics is well known. Any divergence from this physics will originate in the hidden correlations. This is why we employ a system of $n$ HOs here.

\vskip 3mm
\nd Divergences constitute an important theme of   theoretical physics.
The study and elimination of these divergences may be one of the most relevant tasks of theoretical
endeavor. The typical example is the (thus far failed, alas)  attempt to
quantify the gravitational field. Examples of  divergences-elimination can be found
in  references \cite{tq1,tq2,tq3,tq4,tq5}.

\nd  We use here  an quite  simplified version (see \cite{tr1}), of the
methodology of \cite{tq1,tq2,tq3,tq4,tq5} with regards to   Tsallis
statistics \cite{tsallis,web}, focusing on  its
applicability to self-gravitation \cite{PP93,chava,lb}.
Divergence's removal will be seen to yield quite interesting insights.

\nd These emerge using mathematics
well known for the last  40 years ago. Their  development
allowed M. Veltman and G. t'Hooft to be  awarded with the
Nobel prize of physics in 1999.
Comfortable acquaintance with these
mathematics is not a prerequisite to follow this paper.
However, one must  accept that their physical significance
 is not now to kin doubt. In fact, one just needs i) analytical
extensions and ii) dimensional regularization \cite{tq1,tq2,tq3,tq4,tq5}.

\nd We will here analyze the behavior of
$Z$ and $<U>$ in connection with  three zones of possible arguments of the
$\Gamma$-function thay appears in $Z$ and $<U>$.
These arguments of the $\Gamma$-function
rule the   $Z$ - $<U>$ behavior, that in turn produces
 three distinct zones,
for a given spatial dimension $\nu$, Tsallis' index
$q$ and number of particles $N$.
The zone's specifics are:\\
$(1)\;\;\;\frac {1} {1-q}-n\nu-1>0$\\
$(2)\;\;\;\frac {1} {1-q}-n\nu<0\;\;\;
\Gamma\left(\frac {1} {1-q}-n\nu\right)>0$\\
$(3)\;\;\;\frac {1} {1-q}-n\nu=-p
\;\;\;p=0,1,2,3,4.....$\\
Normal behavior is
found in zone (1). Something resembling what might constitute gravitational
effects (GE) are encountered in zone (2).
In zone (3) we find  both normal
behavior and also GE (Also known as gravotermal effects).

\nd {\bf Remark than in instance (3) we are performing a
regularization of the corresponding theory, not
a renormalization.}

\section{The Harmonic Oscillator}

\setcounter{equation}{0}
It has to be noted, from the beginning,  that   we use in this contribution
normal (linear in the probability) expectation values. For simplicity reasons, we do not appeal to the weighted
ones, customarily attached to Tsallis-related papers \cite{tsallis}. In this case one  restricts oneself to
the interval $[0< q \le 1]$, and, consequently, the so-called Tsallis cut-off  problem \cite{tsallis} is avoided.

For the q-partition function one has
\[Z=V^n\int\limits_{-\infty}^{\infty}
\left[1+\beta(1-q)(p_1^2+\cdot\cdot\cdot
p_n^2+q_1^2+\cdot\cdot\cdot q_n^2)
\right]^{\frac {1}{q-1}}\otimes\]
\begin{equation}
\label{eq1.1}
d^{\nu}p_1\cdot\cdot\cdot d^{\nu}p_n
d^{\nu}q_1\cdot\cdot\cdot d^{\nu}q_n,
\end{equation}
Or
\begin{equation}
\label{eq1.2}
Z=\frac {2\pi^{\nu n}}
{\Gamma\left(\nu n\right)}
\int\limits_{-\infty}^{\infty}
\left[1+\beta(1-q) p^2
\right]^{\frac {1}{q-1}}p^{2\nu n-1}dp.
\end{equation}
We have integrated over the angles and taken
$p^2=p_1^2+\cdot\cdot\cdot p_n^2+q_1^2+\cdot\cdot\cdot q_n^2$.
Changing variables in the fashion  $x=p^2$, the
last  integral becomes
\begin{equation}
\label{eq1.3}
Z=\frac {\pi^{\nu n}}
{\Gamma\left(\nu n\right)}
\int\limits_{-\infty}^{\infty}
\left[1+\beta(1-q) x
\right]^{\frac {1}{q-1}}x^{\nu n-1}dx,
\end{equation}
that evaluated, yields
\begin{equation}
\label{eq1.4}
Z=\left[\frac {\pi} {\beta(1-q)}\right]^{\nu n}
\frac {\Gamma\left(\frac {1} {1-q}-\nu n\right)}
{\Gamma\left(\frac {1} {1-q}\right)}.
\end{equation}
Similarly we have
\[Z<U>=\int\limits_{-\infty}^{\infty}
\left[1+\beta(1-q)(p_1^2+\cdot\cdot\cdot p_n^2+
q_1^2+\cdot\cdot\cdot q_n^2)\right]^{\frac {1}{q-1}}\]
\begin{equation}
\label{eq1.5}
(p_1^2+\cdot\cdot\cdot p_n^2+q_1^2+\cdot\cdot\cdot q_n^2)
d^{\nu}p_1\cdot\cdot\cdot d^{\nu}q_n.
\end{equation}
In spherical coordinates this becomes
\begin{equation}
\label{eq1.6}
Z<U>=\frac {2\pi^{\nu n}}
{\Gamma\left(\frac {\nu n} {2}\right)}
\int\limits_{-\infty}^{\infty}
\left[1+\beta(1-q) p^2
\right]^{\frac {1}{q-1}}p^{2\nu n+1}dp,
\end{equation}
and setting $x=p^2$
this is now
\begin{equation}
\label{eq1.7}
Z<U>=\frac {\pi^{\nu n}}
{\Gamma\left(\nu n\right)}
\int\limits_{-\infty}^{\infty}
\left[1+\beta(1-q) x
\right]^{\frac {1}{q-1}}x^{\nu n}dx,
\end{equation}
that evaluated yields
\begin{equation}
\label{eq1.8}
<U>=\frac {1} {Z}\frac {\nu n} {\beta(1-q)}
\left[\frac {\pi} {\beta(1-q)}\right]^{\nu n}
\frac {\Gamma\left(\frac {1} {1-q}-\nu n
-1\right)} {\Gamma\left(\frac {1} {1-q}\right)},
\end{equation}
or
\begin{equation}
\label{eq1.9}
<U>=\frac {\nu n} {\beta[q-\nu n(1-q)]}.
\end{equation}
The derivative with respect to  $T$ yields for the specific heat $C$
at constant  volume
\begin{equation}
\label{eq1.10}
C=\frac {\nu nk} {q-\nu n(1-q)}.
\end{equation}

\section{Limitations that restrict the particle-number}

\setcounter{equation}{0}

We saw in  Ref. \cite{tr1}, for an ideal q-gas, that its number of particles $N$ becomes restricted due to hidden q-correlations. Some related work by  Livadiotis,  McComas, and Obregon,  should be mentioned
\cite{li1,li2,obreg}. \vskip 3mm \nd

Our original  presentation begins here. We detect a similar effect below for our system of $n$ classical HOs.  We analyze first the Gamma functions
involved in evaluating  $Z$ and $<U>$, for the zone $[0 < q \le 1]$.
Starting from  (\ref{eq2.1}) we get, for a positive Gamma-argument
\begin{equation}
\label{eq2.1}
\frac {1} {1-q}-\nu n>0.
\end{equation}
In analogous fashion we have from  (\ref{eq2.2})
\begin{equation}
\label{eq2.2}
\frac {1} {1-q}-\nu n-1>0.
\end{equation}
We are confronted  then with  two conditions that  strictly  limit the
particle-number $n$, that is,
\begin{equation}
\label{eq2.3}
1\leq n<\frac {q} {\nu(1-q)}
\end{equation}
There is a maximum allowable $n$.
For instance, if
$q=1-10^{-3}, \nu=3$, we have
\begin{equation}
\label{eq2.4}
1\leq n<333,
\end{equation}
and one can not exceed 332 particles.
\section{The dimensional  analytical extension of divergent integrals \cite{tq1,tq2,tq3,tq4,tq5}}
\setcounter{equation}{0}

We study first  negative Gamma arguments in (\ref{eq1.4}). They will demand analytical extension/dimensional regularization of the integrals (1.4) and (1.8). Accordingly,

\begin{equation}
\label{eq3.1}
\frac {1} {1-q}-\nu n<0,
\end{equation}
together with
\begin{equation}
\label{eq3.2}
\Gamma\left(\frac {1} {1-q}-\nu n\right)>0.
\end{equation}
Utilize now
\begin{equation}
\label{eq3.3}
\Gamma(z)\Gamma(1-z)=\frac {\pi} {\sin(\pi z)},
\end{equation}
to encounter

\begin{equation}
\label{eq3.4}
\Gamma\left(\frac {1} {1-q}-\nu n\right)=-
\frac {\pi} {\sin\pi\left(\nu n-\frac {1} {1-q}\right)
\Gamma\left(\nu n+1-\frac {1} {1-q}\right)}>0.
\end{equation}
The above is true if
\begin{equation}
\label{eq3.5}
\sin\pi\left(\nu n-\frac {1} {1-q}\right)<0,
\end{equation}
so that
\begin{equation}
\label{eq3.6}
2p+1<\nu n-\frac {1} {1-q}<2(p+1)
\end{equation}
where $p=0,1,2,3,4,5.....$, or equivalently

\begin{equation}
\label{eq3.9}
\frac {2p+1} {\nu}+\frac {1} {\nu(1-q)}<n<
\frac {2(p+1)} {\nu}+\frac {1} {\nu(1-q)}.
\end{equation}
We note that, from  (\ref{eq3.1}), (\ref{eq3.2}), and  (\ref{eq3.4})
we find (1) $Z>0$, (2) $<U><0$ (Einstein crystal), (3) $C<0$,
which entails bound states, on account of (2) and self-gravitation according to (3) \cite{lb}.

\section{The poles of the Harmonic Oscillator treatment}

\setcounter{equation}{0}

If the Gamma's argument is such that

\begin{equation}
\label{eq4.1}
\frac {1} {1-q}-\nu n=-p\;\;{\rm for} \;\;p=0,1,2,3,......,
\end{equation}
 $Z$ exhibits a single pole. \vskip 3mm
\nd For $\nu=1$ one has
\begin{equation}
\label{eq4.2}
\frac {1} {1-q}-n=-p\;\;{\rm for} \;\;p=0,1,2,3,.......
\end{equation}
\nd Given that   $0\leq q<1$, the pertinent
 $q$ values become
\begin{equation}
\label{eq4.3}
q=\frac {1} {2},\frac {2} {3},\frac {3} {4},\frac {4} {5},......,
\end{equation}
$n\geq 2$. \vskip 3mm
\nd For $\nu=2$
\begin{equation}
\label{eq4.4}
\frac {1} {1-q}-2n=-p\;\;{\rm for} \;\;p=0,1,2,3,......,
\end{equation}
\nd Once more, since  $0\leq q<1$,
\begin{equation}
\label{eq4.5}
q=\frac {1} {2},\frac {2} {3},\frac {3} {4},\frac {4} {5},......,
\end{equation}
$n\geq 1$.
\vskip 3mm
\nd For $\nu=3$
\begin{equation}
\label{eq4.6}
\frac {1} {1-q}-3n=-p\;\;{\rm for} \;\;p=0,1,2,3,......,
\end{equation}
and since  $0\leq q<1$,
\begin{equation}
\label{eq4.7}
q=\frac {1} {2},\frac {2} {3},\frac {3} {4},\frac {4} {5},......,
\end{equation}
$n\geq 1$. \vskip 3mm

\nd We tackle now poles in $<U>$. They result from

\begin{equation}
\label{eq4.8}
\frac {1} {1-q}-\nu n-1=-p\;\;{\rm for} \;\;p=0,1,2,3,......,
\end{equation}
for $\nu=1$.
\begin{equation}
\label{eq4.9}
\frac {1} {1-q}-n-1=-p\;\;{\rm for} \;\;p=0,1,2,3,......,
\end{equation}
Since $0\leq q<1$, one has
\begin{equation}
\label{eq4.10}
q=\frac {1} {2},\frac {2} {3},\frac {3} {4},\frac {4} {5},......,
\end{equation}
\nd for $\nu=2$.
\begin{equation}
\label{eq4.11}
\frac {1} {1-q}-2n-1=-p\;\;{\rm for} \;\;p=0,1,2,3,......,
\end{equation}
\begin{equation}
\label{eq4.12}
q=\frac {1} {2},\frac {2} {3},\frac {3} {4},\frac {4} {5},......,
\end{equation}
\nd For $\nu=3$
\begin{equation}
\label{eq4.13}
\frac {1} {1-q}-3n-1=-p\;\;{\rm for} \;\;p=0,1,2,3,......,
\end{equation}

\begin{equation}
\label{eq4.14}
q=\frac {1} {2},\frac {2} {3},\frac {3} {4},\frac {4} {5},......,
\end{equation}

\section{The three-dimensional scenario}

\setcounter{equation}{0}

As an illustration of dimensional regularization  \cite{tq1,tq2,tq3,tq4,tq5}
we discuss into some detail the dealing with  the poles at  $q=\frac {1} {2}$ and $q=\frac {2} {3}$.

\subsection{Pole at $q=1/2$}

One has
\begin{equation}
\label{eq5.1}
Z=\left(\frac {2\pi} {\beta}\right)^{\nu n}
\Gamma\left(2-\nu n\right).
\end{equation}
Using
\begin{equation}
\label{eq5.2}
\Gamma\left(2-\nu n\right)
\Gamma\left(\nu n-1\right)=
-\frac {\pi} {\sin\left(\pi\nu n\right)}
\end{equation}
or, equivalently

\begin{equation}
\label{eq5.3}
\Gamma\left(2-\nu n\right)
\Gamma\left(\nu n-1\right)=
\frac {(-1)^{3n+1}\pi} {\sin\left[
\pi n (\nu-3)\right]},
\end{equation}
so that
\begin{equation}
\label{eq5.4}
Z=\left(\frac {2\pi} {\beta}\right)^{\nu n}
\frac {(-1)^{3n+1}\pi}
{\sin[\pi n(\nu-3)]
\Gamma\left(\nu n-1\right)}.
\end{equation}
Given that
\begin{equation}
\label{eq5.5}
\sin[\pi n(\nu-3)]=\pi n(\nu-3)
\left\{1+\sum\limits_{m=1}^{\infty}
\frac {(-1)^m} {(2m+1)!} \left[\pi n
(\nu-3)\right]^{2m}\right\}=
\end{equation}
\begin{equation}
\label{eq5.6}
=\pi n(\nu-3)X,
\end{equation}
with
\begin{equation}
\label{eq5.7}
X=
\left\{1+\sum\limits_{m=1}^{\infty}
\frac {(-1)^m} {(2m+1)!} \left[\pi n
(\nu-3)\right]^{2m}\right\},
\end{equation}
we obtain
\begin{equation}
\label{eq5.8}
Z=\left(\frac {2\pi} {\beta}\right)^{3 n}
\frac {(-1)^{n+1}}
{\Gamma\left(\nu n-1\right)
Xn(\nu-3)}
\left[1+n(\nu-3)\ln\left(\frac {2\pi} {\beta}\right)+
\cdot\cdot\cdot\right]
\end{equation}
The term independent of  $\nu-3$ is, according to dimensional regularization recipes
\cite{tq1,tq2,tq3,tq4,tq5}
\begin{equation}
\label{eq5.9}
Z=\left(\frac {2\pi} {\beta}\right)^{3 n}
\frac {(-1)^{n+1}}
{\Gamma\left(3 n-1\right)}
\ln\left(\frac {2\pi} {\beta}\right)
\end{equation}
This $Z$  is then the physical one at the pole \cite{tq1,tq2,tq3,tq4,tq5}. Now, for the mean energy one has

\begin{equation}
\label{eq5.10}
Z<U>=\frac {2n\nu} {\beta}
\left(\frac {2\pi} {\beta}\right)^{\nu n}
\Gamma\left(1-\nu n\right).
\end{equation}
Employing
\begin{equation}
\label{eq5.11}
\Gamma\left(1-\nu n\right)
\Gamma\left(\nu n\right)=
\frac {\pi} {\sin\left(\pi\nu n\right)}
\end{equation}
or, equivalently
\begin{equation}
\label{eq5.12}
\Gamma\left(1-\nu n\right)
\Gamma\left(\nu n\right)=
\frac {(-1)^{3n}\pi} {\sin\left[
\pi n (\nu-3)\right]}
\end{equation}
we encounter for  $<U>$

\begin{equation}
\label{eq5.13}
Z<U>=\frac {2n\nu} {\beta}
\left(\frac {2\pi} {\beta}\right)^{\nu n}
\frac {(-1)^{3n}\pi}
{\sin[\pi n(\nu-3)]
\Gamma\left(\nu n\right)}.
\end{equation}
$<U>$ can be rewritten in the fashion

\[Z<U>=\frac {n(\nu-3)} {\beta}
\left(\frac {2\pi} {\beta}\right)^{\nu n}
\frac {(-1)^{3n}\pi}
{\sin[\pi n(\nu-3)]
\Gamma\left(\nu n\right)} +\]
\begin{equation}
\label{eq5.14}
\frac {6n} {\beta}
\left(\frac {2\pi} {\beta}\right)^{3 n}
\frac {(-1)^{3n}\pi}
{\sin[\pi n(\nu-3)]
\Gamma\left(\nu n\right)}.
\end{equation}
 Recalling the Z-procedure gives  for  $<U>$

\[Z<U>=\frac {2} {\beta}
\left(\frac {2\pi} {\beta}\right)^{3 n}
\frac {(-1)^{3n}}
{\Gamma\left(3 n\right)}+\]
\begin{equation}
\label{eq5.15}
\frac {6n} {\beta}
\left(\frac {2\pi} {\beta}\right)^{3 n}
\frac {(-1)^{3n}}
{\Gamma\left(3 n\right)}
\ln\left(\frac {2\pi} {\beta}\right)
\end{equation}
or,  equivalently

\begin{equation}
\label{eq5.16}
Z<U>=\frac {2} {\beta}
\left(\frac {2\pi} {\beta}\right)^{3n}
\frac {(-1)^{3n}}
{\Gamma\left(3 n\right)}
\left[1+3n
\ln\left(\frac {2\pi} {\beta}\right)\right].
\end{equation}
Remembering now  (\ref{eq5.9}) for the physical $Z$
 on arrives at

\begin{equation}
\label{eq5.17}
<{\cal U}>=
\frac {2} {\beta(3n-1)}
\left[\frac {1} {\ln\beta-\ln 2\pi}-
3n\right].
\end{equation}
one treats first $(-1)^{3n+1}=-1$, so that   $n=2,4,6,8,......$, and

\begin{equation}
\label{eq5.18}
Z=\frac {1}
{\Gamma\left(3 n-1\right)}
\left(\frac {2\pi} {\beta}\right)^{3 n}
\ln\left(\frac {\beta} {2\pi}\right)
\end{equation}
If
$(-1)^{3n+1}=1$, then
$n=1,3,5,7......$ and
\begin{equation}
\label{eq5.19}
Z=\frac {1}
{\Gamma\left(3 n-1\right)}
\left(\frac {2\pi} {\beta}\right)^{3 n}
\ln\left(\frac {2\pi} {\beta}\right).
\end{equation}
According to  (\ref{eq5.17}) - (\ref{eq5.18}) and asking
$Z>0$ and  $<U>>0$ one finds
\begin{equation}
\label{eq5.20}
\frac {1} {2\pi ke^{\frac {1} {3n}}}<T<
\frac {1} {2\pi k}.
\end{equation}
From  (\ref{eq5.17}) -  (\ref{eq5.19}) and requiring
$Z>0$ y $<U><0$ (Einstein crystal) one encounters
\begin{equation}
\label{eq5.21}
0\leq T<\frac {1} {2\pi ke^{\frac {1} {3n}}}
\end{equation}
The specific heat is derived  from
(\ref{eq5.17}) for $<U>$. We have
\begin{equation}
\label{eq5.22}
C=
\frac {2k} {3n-1}
\left[\frac {1} {\ln\beta-\ln 2\pi}+
\frac {1} {(\ln\beta-\ln 2\pi)^2}-
3n\right].
\end{equation}
$C$ depends on $T$ and this is a quantum effect, since classically $C$ is a constant. Also, $C$ depends on $T$ because of the excitation of internal degrees of freedom, which the poles somehow detect.

\subsection{The Pole at $q=2/3$}

Now  $Z$ is
\begin{equation}
\label{eq5.23}
Z=\left(\frac {3\pi} {\beta}\right)^{\nu n}
\frac {\Gamma\left(3-\nu n\right)}
{\Gamma\left(3\right)}.
\end{equation}
Employing once  again
\begin{equation}
\label{eq5.24}
\Gamma\left(3-\nu n\right)
\Gamma\left(\nu n-2\right)=
\frac {\pi} {\sin\left(\pi\nu n\right)},
\end{equation}
or, equivalently

\begin{equation}
\label{eq5.25}
\Gamma\left(3-\nu n\right)
\Gamma\left(\nu n-2\right)=
\frac {(-1)^{3n}\pi} {\sin\left[
\pi n (\nu-3)\right]},
\end{equation}
so that we have
\begin{equation}
\label{eq5.26}
Z=\frac {1} {2}
\left(\frac {3\pi} {\beta}\right)^{\frac {\nu n} {2}}
\frac {(-1)^{3n}\pi}
{\sin[\pi n(\nu-3)]
\Gamma\left(\nu n-1\right)}.
\end{equation}
One then  dimensionally regularizes  $Z$ - $<U>$ as done for the previous pole, to reach

\begin{equation}
\label{eq5.27}
Z=\frac {1} {2}
\left(\frac {3\pi} {\beta}\right)^{3 n}
\frac {(-1)^{3n}}
{\Gamma\left(3 n-2\right)}
\ln\left(\frac {3\pi} {\beta}\right),
\end{equation}
\begin{equation}
\label{eq5.28}
<{\cal U}>=
\frac {3} {2\beta(3n-2)}
\left[\frac {1} {\ln\beta-\ln 3\pi}-
3n\right].
\end{equation}
We seal  first with
$(-1)^{\frac {3n-1} {2}}=-1$ and then
$n=1,3,5,7,9......$, so that
\begin{equation}
\label{eq5.29}
Z=\frac {1}
{2\Gamma\left(3n-2\right)}
\left(\frac {3\pi} {\beta}\right)^{3 n}
\ln\left(\frac {\beta} {3\pi}\right).
\end{equation}
For
$(-1)^{3n}=1$, one has
$n=2,4,6,8......$ and
\begin{equation}
\label{eq5.30}
Z=\frac {1}
{2\Gamma\left(3n-2\right)}
\left(\frac {3\pi} {\beta}\right)^{3 n}
\ln\left(\frac {3\pi} {\beta}\right).
\end{equation}

According to   (\ref{eq5.28}) - (\ref{eq5.29}) and asking

$Z>0$ - $<U>>0$ we encounter
\begin{equation}
\label{eq5.31}
\frac {1} {3\pi ke^{\frac {1} {3n}}}<T<
\frac {1} {3\pi k}.
\end{equation}
From  (\ref{eq5.28})-y (\ref{eq5.30}) and asking
$Z>0$ - $<U><0$ we obtain

\begin{equation}
\label{eq5.32}
0\leq T<\frac {1} {3\pi ke^{\frac {1} {3n}}}.
\end{equation}
As for  $C$ we have
\begin{equation}
\label{eq5.33}
C=
\frac {3k} {2(3n-2)}
\left[\frac {1} {\ln\beta-\ln 3\pi}+
\frac {1} {(\ln\beta-\ln 3\pi)^2}-
3n\right].
\end{equation}

\section{Conclusions}


\setcounter{equation}{0}

\nd Here one has appealed to  an elementary
regularization method to study  the poles in both thae partition function $Z$ and
the mean energy  $<U>$ for particular, discrete  values of Tsallis' parameter q
in a non additive q-scenario. After investigating  the thermal
behavior at the poles, we found interesting features, like what might possibly constitute  self-gravitation or quantum effects. The
analysis was made for one, two, three, and $N$ dimensions. We discover
 pole-characteristics that are  unexpected but  true. In particular:

\begin{itemize}

\item An upper bound to the
temperature at the poles, in agreement with the findings  of Ref.
\cite{PP94}.

\item In some circumstances, Tsallis' entropies are positive only for a
restricted temperature-range.

\item Negative specific heats, which might constitute signatures  of
self-gravitating systems \cite{lb}, are encountered.

\item If the system is bound, we can regard it as a  "classical" Einstein-crystal. But we have
for it a  temperature dependence of the specific
heat.

\item Thus, we find at the poles, that i) the specific heat  $C$ is temperature ($T$) dependent (classically!), and, ii) self-gravitational effects. Note that $C$ can become $T-$dependent only due to internal degrees
of freedom, and that this is a quantum effect. We detect this dependence  here at a purely classical level.

\end{itemize}

\nd These  physical results are collected employing just statistical consideration,  not mechanical ones. This  might perhaps  remind one of  a similar
feature associated to  the entropic force conjectured by
Verlinde \cite{verlinde}.

\nd The Tsalllis' rule

$$S_q(A,B) = S_q(A) + S_q(B) + (1-q) S_q(A)S:q(B); \,\,\, q \in {\cal R},$$
is seen here to erect a far from trivial scenario, in which strange effects take place.

\newpage
\begin{figure}[h]
\begin{center}
\includegraphics[scale=0.6,angle=0]{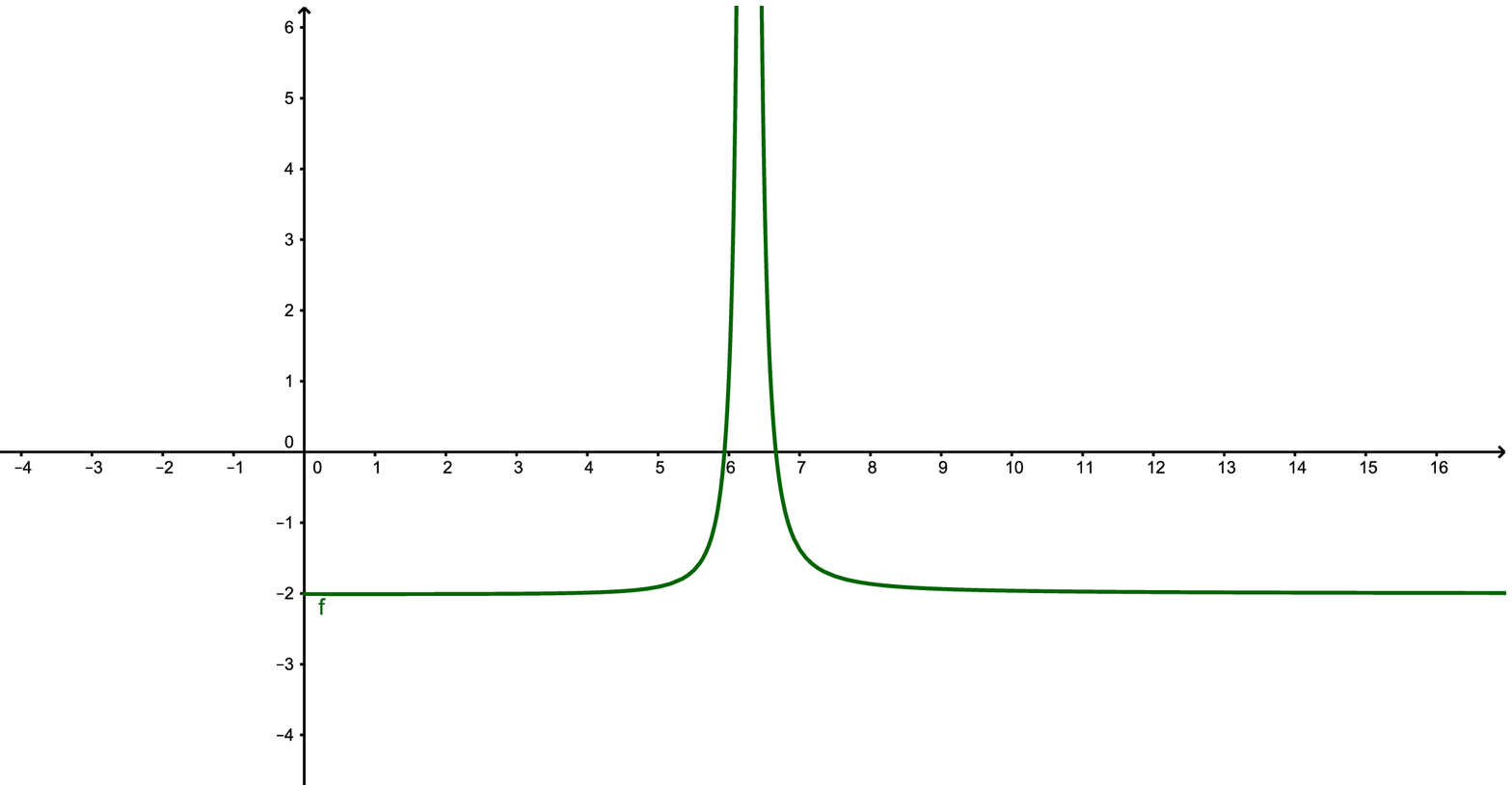}
\vspace{-0.2cm} \caption{$c/k$ at the pole $q=1/2$
versus $\beta$ with $n=100$.
The right branch corresponds to $Z>0$, i.e.,
the physical branch.}\label{fig1}
\end{center}
\end{figure}

\newpage
\begin{figure}[h]
\begin{center}
\includegraphics[scale=0.6,angle=0]{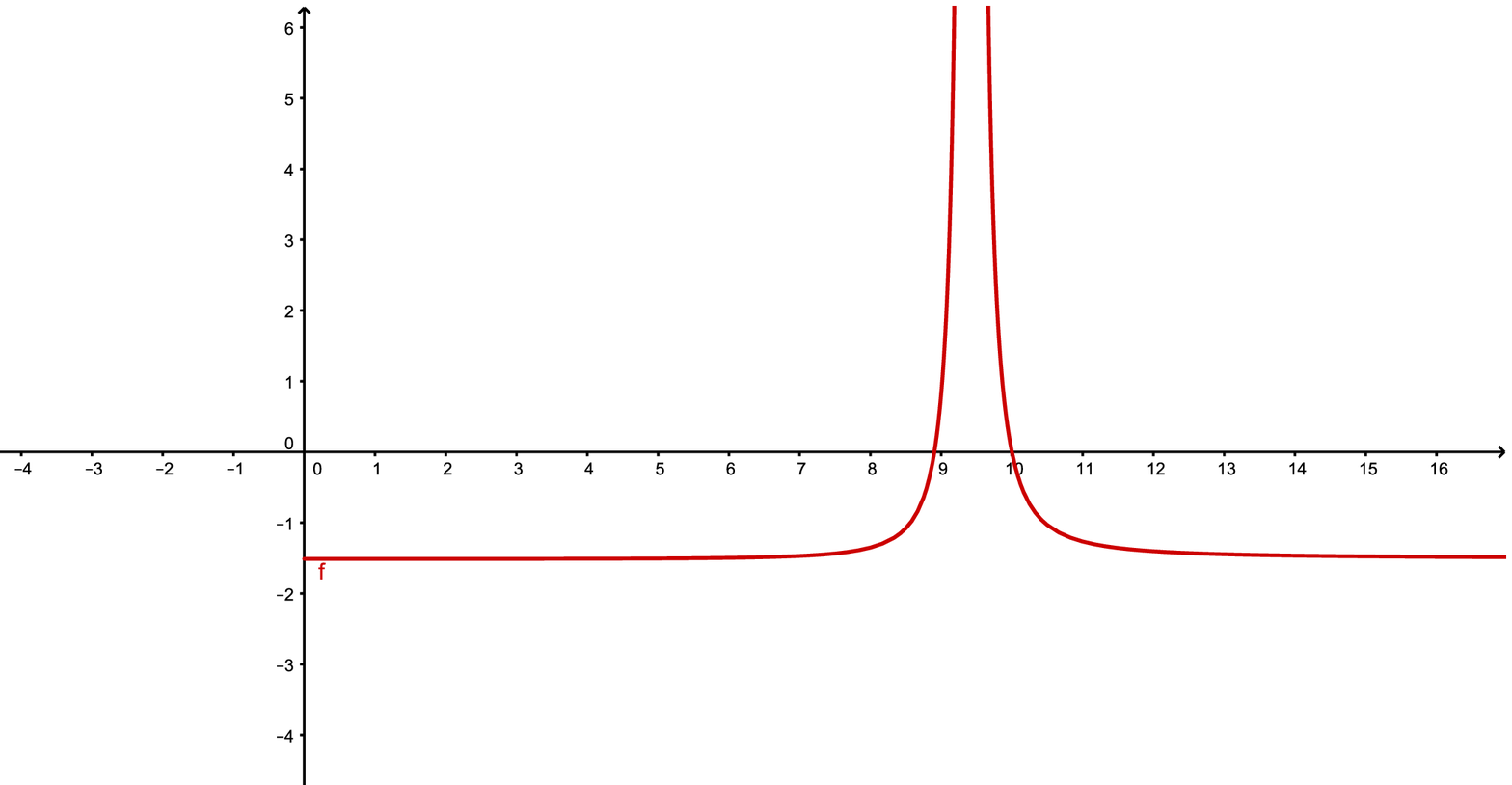}
\vspace{-0.2cm} \caption{$c/k$ at the pole $q=2/3$
versus $\beta$, with $n=99$.
The right branch corresponds to $Z>0$, i.e.,
the physical branch.}\label{fig2}
\end{center}
\end{figure}


\newpage

\begin{thebibliography}{99}

\bibitem{tsallis}  M. Gell-Mann and C. Tsallis, Eds. {\it Nonextensive Entropy:
Interdisciplinary applications}, Oxford University Press, Oxford,
2004;  C. Tsallis, {\it Introduction to Nonextensive Statistical
Mechanics: Approaching a Complex World}, Springer, New York, 2009.

\bibitem{web} See http://tsallis.cat.cbpf.br/biblio.htm for a
regularly updated bibliography on the subject.

\bibitem{pre} P-Jizba, J. Korbel,  V. Zatloukal,
Phys. Rev. E {\bf 95}, 022103 (2017); G. B. Bagci,  T.  Oikonomou,
Phys. Rev. E {\bf 93}, 022112 (2016); L. S. F. Olavo,
Phys. Rev. E  {\bf 64}, 036125 (2001).


\bibitem{epjb1} I. S. Oliveira:
Eur. Phys. J. B {\bf 14}, 43 (2000)
\bibitem{epjb2} E. K. Lenzi , R. S. Mendes:
Eur. Phys. J. B {\bf 21}, 401 (2001)
\bibitem{epjb3} C. Tsallis:
Eur. Phys. J. A {\bf 40}, 257 (2009)
\bibitem{epjb4} P. H. Chavanis:
Eur. Phys. J. B {\bf 53}, 487 (2003)
\bibitem{epjb5} G. Ruiz ,C. Tsallis:
Eur. Phys. J. B {\bf 67}, 577 (2009)
\bibitem{epjb6} P. H. Chavanis , A. Campa:
Eur. Phys. J. B {\bf 76}, 581 (2010)
\bibitem{epjb7} N. Kalogeropoulos:
Eur. Phys. J. B {\bf 87}, 56 (2014)
\bibitem{epjb8} N. Kalogeropoulos:
Eur. Phys. J. B {\bf 87}, 138 (2014)
\bibitem{epjb9} A. Kononovicius, J. Ruseckas:
Eur. Phys. J. B {\bf 87}, 169 (2014)

\bibitem{PP93} A. R. Plastino, A. Plastino,  Phys.  Lett. A {\bf 174} (1993)
384.

\bibitem{chava} P. H. Chavanis, C. Sire, Physica A {\bf 356} (2005)
419; P.-H. Chavanis, J. Sommeria, Mon. Not. R. Astron. Soc. {\bf
296} (1998) 569.


\bibitem{lb} D. Lynden-Bell,  R. M. Lynden-Bell,  Mon. Not. R. Astron. Soc. {\bf
181} (1977) 405.



\bibitem{tp11o} C. Tsallis, {\it Introduction to Nonextensive Statistical Mechanics}
(Springer, Berlin, 2009).

\bibitem{tp11}   F. Barile et al. (ALICE Collaboration),
EPJ Web Conferences {\bf 60}, (2013) 13012; B. Abelev et al.
(ALICE Collaboration), Phys. Rev. Lett. {\bf 111}, (2013) 222301;
Yu. V.Kharlov (ALICE Collaboration), Physics of Atomic Nuclei {\bf
76}, (2013) 1497. ALICE Collaboration,  Phys. Rev. C 91, (2015)
024609; ATLAS Collaboration, New J. Physics {\bf 13}, (2011)
053033; CMS Collaboration, J. High Energy Phys. {\bf 05}, (2011)
064; CMS Collaboration, Eur. Phys. J. C {\bf 74}, (2014) 2847.

\bibitem{phenix}  A. Adare et al (PHENIX Collaboration), Phys. Rev.
D {\bf   83}, (2011) 052004;  PHENIX Collaboration,  Phys. Rev. C
{\bf   83}, (2011) 024909; PHENIX Collaboration, Phys. Rev. C
{\bf   83}, (2011) 064903; PHENIX Collaboration,  Phys. Rev. C
{\bf   84}, (2011) 044902.




\bibitem{epjb} A. Plastino; M. C. Rocca; G. L. Ferri:
EPJB {\bf 89}, 150 (2016).
\bibitem{arxiv} A. Plastino, M. C. Rocca:
arXiv:1702.03535 (2017).
\bibitem{tq1} C. G. Bollini and J. J. Giambiagi: Phys. Lett.
{\bf B 40}, (1972), 566.Il Nuovo Cim. {\bf B 12}, (1972), 20.
\bibitem{tq2} C. G. Bollini and J.J Giambiagi :
Phys. Rev. {\bf D 53}, (1996), 5761.
\bibitem{tq3} G.'t Hooft and M. Veltman: Nucl. Phys. {\bf B 44},
(1972), 189.
\bibitem{tq4} C. G. Bollini, T. Escobar and M. C. Rocca :
Int. J. of Theor. Phys. {\bf 38}, (1999), 2315.
\bibitem{tq5} C. G Bollini and M.C. Rocca :
Int. J. of Theor. Phys. {\bf 43}, (2004), 59. Int. J. of Theor.
Phys. {\bf 43}, (2004), 1019. Int. J. of Theor. Phys. {\bf 46},
(2007), 3030.


\bibitem{tr1} A. Plastino, M. C. Rocca: arXiv:1701.03525

\bibitem{li1} G. Livadiotis, D. J. McComas:
APJ {\bf 748}, 88 (2011).
\bibitem{li2} G. Livadiotis: Entropy
{\bf 17}, 2062 (2015).

\bibitem{obreg} O. Obregon, International Journal of Modern Physics A
{\bf 30} (2015) 1530039.



\bibitem{PP94} A. R. Plastino, A. Plastino, Phys. Lett. A  {\bf 193}  (1994) 140.








\bibitem{verlinde} E. Verlinde,  arXiv:1001.0785 [hep-th];
JHEP 04, 29 (2011).

\end{thebibliography}
\end{document}